\documentclass{vgtc}     

\graphicspath{{figures/}{pictures/}{images/}{./}} 

\usepackage{times}   
\usepackage{soul}
\usepackage[dvipsnames,svgnames]{xcolor}
\usepackage{subcaption}
\usepackage{diagbox}  
\usepackage{array}    
\usepackage{color,soul}
\usepackage{sidecap}
\usepackage{tikz}
\usepackage{fancyhdr}
\usepackage{lipsum} 
\usepackage{tikz}

\usepackage{tabu}                      
\usepackage{booktabs}                  
\usepackage{lipsum}                    
\usepackage{mwe}                       
\usepackage{mathptmx}                  

\onlineid{2558}
\vgtccategory{Research}
\vgtcinsertpkg

\title{DataliVR: Transformation of Data Literacy Education through Virtual Reality with ChatGPT-Powered Enhancements}

\author{Hong Gao\thanks{e-mail: hong.gao@tum.de}\\ %
        \scriptsize Human-Centered Technologies for Learning\\ 
        \scriptsize Technical University of Munich 
\and Haochuan Huai \\
\scriptsize Technical University of Munich 
\and Sena Yildiz-Degirmenci \\
\scriptsize Teaching and Learning with Digital Media \\
\scriptsize Technical University of Munich
\and Maria Bannert \\
\scriptsize Teaching and Learning with Digital Media \\
\scriptsize Technical University of Munich
\and Enkelejda Kasneci \\ %
\scriptsize Human-Centered Technologies for Learning\\
\scriptsize Technical University of Munich %
}

\abstract{
Data literacy is essential in today's data-driven world, emphasizing individuals' abilities to effectively manage data and extract meaningful insights. However, traditional classroom-based educational approaches often struggle to fully address the multifaceted nature of data literacy. As education undergoes digital transformation, innovative technologies such as Virtual Reality (VR) offer promising avenues for immersive and engaging learning experiences. This paper introduces \textbf{DataliVR}, a pioneering VR application aimed at enhancing the data literacy skills of university students within a contextual and gamified virtual learning environment. By integrating Large Language Models (LLMs) like \textbf{ChatGPT} as a conversational artificial intelligence (AI) chatbot embodied within a virtual avatar, DataliVR provides personalized learning assistance, enriching user learning experiences. Our study employed an experimental approach, with chatbot availability as the independent variable, analyzing learning experiences and outcomes as dependent variables with a sample of thirty participants. Our approach underscores the effectiveness and user-friendliness of ChatGPT-powered DataliVR in fostering data literacy skills. Moreover, our study examines the impact of the ChatGPT-based AI chatbot on users' learning, revealing significant effects on both learning experiences and outcomes. Our study presents a robust tool for fostering data literacy skills, contributing significantly to the digital advancement of data literacy education through cutting-edge VR and AI technologies. Moreover, our research provides valuable insights and implications for future research endeavors aiming to integrate LLMs (e.g., ChatGPT) into educational VR platforms.
} %


\begin{document}

\maketitle


\vfill 
\noindent\footnotesize \copyright 2024 IEEE. Personal use of this material is permitted. Permission from IEEE must be obtained for all other uses, in any current or future media, including reprinting/republishing this material for advertising or promotional purposes, creating new collective works, for resale or redistribution to servers or lists, or reuse of any copyrighted component of this work in other works. \textit{This is the author's version of an article that has been accepted for publication in ISMAR 2024. The final version will be available online at IEEE Xplore. DOI will be assigned once published.}

\vspace{1ex} 

\section{Introduction} 

In the age of digital transformation, data literacy has emerged as an essential 21st-century skill, defined as ``the ability to understand, interpret, and effectively use data to make informed decisions''~\cite{datalidef2016}. Recently, the significance of this skill has transcended the confines of computer science and has gained prominence in various scientific disciplines where data holds significant importance. Professionals such as researchers and data librarians rely on data literacy to proficiently manage and analyze datasets~\cite{dl_reandli17}. Similarly, education professionals need to impart and apply this competency in their teaching methods and professional practices~\cite{dledupro21, dleduhigher}. Students, especially in higher education, should develop and apply data literacy in their academic endeavors to adequately prepare for future careers and lifelong learning in today's data-centric society. Recognizing the significance of data literacy, it becomes imperative to address the existing challenges related to the practical and effective development and assessment of individuals' data literacy. 

Data literacy skills are intrinsically linked to data management practices~\cite{datamange18}, emphasizing ``the ability to handle data throughout its lifecycle, including collection, management, evaluation, and data application in a critical manner''~\cite{dldef2015}. Thus, enhancing data literacy can be achieved through targeted training in data management practices. Despite increasing recognition and the incorporation of data literacy into educational curricula spanning from primary schools to higher education~\cite{dlhighereducation, dlprogram13, wolff16DLurban}, its practical implementation and effectiveness remain limited. Conventional teaching methods, such as lectures and educational programs, offer students hands-on opportunities to work with data~\cite{wolff16DLurban}; however, they may not always effectively engage students. Moreover, the assessment of data literacy is essential for adaptive skill development but is often overlooked and lacks comprehensive research. Current assessment tools mainly rely on traditional methods like tests and self-reported questionnaires, which may not accurately measure students' performance-based data literacy competencies~\cite{dlassessment}. Given these research limitations, there is an immediate need for innovative and practical training and assessment tools for data literacy. 

Virtual Reality (VR) has gained significant popularity in recent years, emerging as a transformative technology with broad applications across various domains, from entertainment and gaming~\cite{vrart22, vredugame20} to healthcare~\cite{vrmedical20} and education~\cite{rVRedu2020, rVR_skilltrain21}. Especially in education, VR stands out as a transformative tool that enhances education by immersing learners in immersive environments, boosting engagement, and improving learning outcomes~\cite{vreduper24}. VR offers several notable advantages. VR provides cost-effective solutions for replicating expensive or hard-to-access learning materials~\cite{hu2021VRchemi} and breaks down geographical limitations by enabling remote learning~\cite{vreduremote22}. Educators can use VR to develop dynamic simulations and interactive lessons tailored to diverse learning needs, improving comprehension and retention of complex concepts often challenging to convey through conventional educational approaches~\cite{vrcheconcept21}. Within the scope of this paper, VR emerges as a potent tool for enhancing data literacy by integrating complex data literacy concepts into immersive and gamified VR environments, addressing a challenge where traditional educational approaches often fall short. 

With rapid advancements in artificial intelligence (AI), large language models (LLMs) have brought about significant changes in various aspects of daily life, including business, healthcare, entertainment, and education~\cite{llms23survey}. Notably, the recent innovation of OpenAI's ChatGPT~\cite{chatgpt} has opened up new possibilities for enhancing educational experiences at all levels of education in a variety of applications~\cite{KASNECI2023102274}, including language learning~\cite{bas2023language}, professional skill development~\cite{chatgptTeacher23}, empowered writing~\cite{KathrinChatgptwri23}, automated essay grading~\cite{chatgptessay23}, and code assistance~\cite{llmscode24}. Particularly noteworthy is the use of ChatGPT-driven conversational agents, which have demonstrated effectiveness in diverse educational settings by offering learners personalized conversational assistance~\cite{survey23chatgptagent, villagptagent23}. While the integration of AI chatbots in VR is not a recent innovation~\cite{vravatarchildren22}, the emergence of ChatGPT-powered chatbots represents a significant advancement, offering enhanced adaptability, nuanced responses, improved contextual understanding, personalized assistance, and a more natural conversational flow compared to traditional AI-driven chatbots. However, despite the increasing integration of ChatGPT-powered chatbots in VR~\cite{genaichat_vr24}, their potential effectiveness and influence on user learning behaviors remain largely unexplored. This research gap motivates our investigation into using a ChatGPT-powered chatbot within a custom-designed VR application aimed at fostering data literacy skills. 

In summary, our contributions can be outlined as follows:
\begin{itemize}
    \item We introduce \textbf{DataliVR}, an innovative VR application aimed at enhancing data literacy skills in higher education by immersing participants in a data lifecycle, with a focus on machine learning classification as the contextual framework for learning. It provides an immersive and gamified learning environment, effectively transforming abstract data literacy concepts into interactive and tangible experiences. The integration of the ChatGPT-powered chatbot further enhances it by offering personalized conversational assistance.
    
    \item We demonstrate the effectiveness and transformative potential of DataliVR through a comprehensive experimental approach, considering various aspects of user learning experiences and outcomes. Our findings highlight its ability to offer users an integrated experience conducive to enhancing data literacy.

    \item We assess the integration of the ChatGPT-powered AI chatbot within DataliVR, examining its effectiveness and impact on user learning experiences and outcomes, offering valuable insights and design considerations for integrating ChatGPT-powered enhancements in educational VR settings. 
\end{itemize} 

\section{Related Work}
\subsection{Data Literacy Education}

Data literacy is increasingly recognized as a crucial competency for individuals navigating the data-centric world. Previous studies have explored various aspects of data literacy education, including teaching methodologies and assessment approaches. For instance, Wolff et al.~\cite{wolff16DLurban} emphasized the importance of data literacy skills by proposing a teaching approach for UK primary schools. Students engaged with interpreting existing visualizations of smart meter data in the context of learning about solar energy generation. Their findings suggested promising prospects for the development of a web-based platform to enhance more complex data skills education in schools. 
Similarly, Simon et al.~\cite{simon22DLcurriculum} introduced a curriculum development model to enhance undergraduate students' data literacy in online astronomy classrooms. A pilot study across nine colleges and universities evaluated the impact of instructional materials on students' beliefs, revealing significant improvements in both science engagement and data literacy skills. Aligned with the learning context of this study, certain curriculum initiatives also aim to enhance digital literacy by incorporating machine learning into the learning objectives for artificial intelligence within secondary education frameworks~\cite{k12aiml23}. Students are expected to describe different approaches to machine learning and explain their basic functionality.
However, existing research has predominantly focused on individual components of data literacy, such as data collection, analysis, interpretation, and visualization. There is a notable absence of comprehensive approaches that systematically teach data literacy skills. Another important aspect of data literacy education is data literacy skills assessment. Various assessment methods have been proposed in previous work~\cite{dlassessment}. For instance, Wu et al.~\cite{wu23dataAnateacher} developed an approach for assessing pre-service teachers' data analysis knowledge based on cognitive diagnostic assessment. Reeves and Chiang~\cite{reevesdlassess19} proposed using a self-reported scale to evaluate educators' data literacy in transforming information into a decision. Oguguo et al.~\cite{oguguo2020assessment} introduced a data literacy questionnaire (DLQ) for assessing the data literacy skills of students from five universities. The DLQ showed effectiveness, and their findings indicated that Ph.D. students exhibited better data literacy skills compared to those at the M.Sc. and B.Sc. levels. Despite the effectiveness of such methods, the existing research on data literacy assessment primarily relies on tests, semi-structured interviews, and questionnaires, which may not fully capture the full spectrum of data literacy competencies. Consequently, there remains a significant gap in the development of comprehensive methodologies that can holistically evaluate integrated data literacy skills. VR holds promise as a potential solution to bridge this gap, enabling integrated and intelligent teaching and evaluation of data literacy.   

\subsection{Virtual Reality in Education}

VR has emerged as a transformative tool in digital education, facilitated by the increasing accessibility of consumer-grade VR headsets, revolutionizing educational experiences across diverse domains. For instance, VR has been used to create immersive simulations for training teachers in professional development~\cite{reviewmrteacher22, Gao_VRteacher23, vrteachertrain21}. Gao et al.~\cite{Gao_VRteacher23} developed an immersive VR classroom tailored for teacher training in classroom management, where teachers delivered virtual lessons to virtual students programmed to display disruptive behaviors. The authors introduced an innovative approach using explainable machine-learning methods based on fused sensor data to assess teachers' expertise levels. Chen~\cite{ChenVRteacherpilot22} demonstrated that immersive VR training significantly enhanced the classroom management skills of pre-service teachers, skills they could effectively transfer to real classroom teaching scenarios. 

Moreover, VR has proven effective in enhancing student learning~\cite{reviewVRStu14, Wu20VRreview}. It offers immersive learning environments that foster improved understanding and retention of intricate concepts in STEM fields such as mathematics~\cite{vrmath23}, physics~\cite{vrphysic21}, science~\cite{vrscience19},  chemistry~\cite{vrchemistry23}, as well as in history~\cite{vrhistoryRoman22} and language learning~\cite{vrlanguage22review}. In addition, VR plays an important role in fostering essential 21st-century core skills, including problem-solving~\cite{ivrproblemsolve21} and critical thinking~\cite{vrcriticalthink20} in computer science education, and soft skills like communication~\cite{vrcomskill20train}. For instance, Pirker et al.~\cite{vrstcsedu20} compared the effectiveness of VR and web-based platforms in sorting algorithm tasks, revealing that immersive visualization and animations in VR significantly improved learning outcomes and positively impacted learners' engagement and emotional experiences. In another study, Segura et al.~\cite{VRockspro20} developed VR-OCKS, a VR game that teaches basic programming concepts to children. This prototype effectively leveraged gamification strategies within VR settings, outperforming traditional 2D environments in motivating children and teenagers to grasp programming skills. Dominic et al.~\cite{vrpairprogram20} explored the potential of VR in supporting pair programming, demonstrating that programmers working in a VR environment exhibited twice the productivity compared to those using conventional screen-sharing systems. Hasenbein et al.~\cite{hasenbein22VRsocial} and Gao et al.~\cite{Gao_VRclass_chi21, gao2023VRgender} have also delved into developing immersive VR classrooms to foster students' computational thinking skills, exploring students' visual perceptions, varied learning experiences, and the interrelationships among them. Their studies addressed research questions that are challenging to explore in real-world classrooms.

Despite the growing body of research on VR's role in education, its potential to support data literacy, a critical 21st-century skill, has been largely overlooked. The multifaceted nature of data literacy, which involves abilities in data management presents unique challenges. Integrating these diverse skills into a coherent and engaging VR-based learning experience proves particularly challenging. These gaps and challenges drive the focus of our study.

\subsection{ChatGPT-powered Education}

The advent of LLMs like ChatGPT (November 2022)~\cite{chatgpt} has brought transformative changes across various domains, significantly impacting education. Particularly, ChatGPT showcases the potential to personalize and enrich learning experiences. For example, Chen et al.~\cite{GPTutor23Chen} introduced GPTutor, a ChatGPT-powered programming tool that delivers concise and accurate code explanations through designed prompts in pop-up messages. Both students and teachers expressed satisfaction with GPTutor for its intuitive interface and effective code clarification. Jauhiainen and Guerra~\cite{suGPTschoolcontent23} evaluated the integration of generative AI (i.e., ChatGPT-3.5) in primary education. They collected feedback from students in grades 4th–6th in two schools regarding their experience with using generative AI-modified learning materials. 
While the results highlighted the considerable potential of ChatGPT in facilitating personalized learning content generation, further analysis based on broader data sets is needed.
Young et al.~\cite{youngChatdialogue23} investigated ChatGPT's potential in generating chatbot dialogues to aid English learning. Their research indicated the effectiveness of the generated chatbot's dialogues in enhancing language acquisition among students. 

Furthermore, ChatGPT has found widespread application for AI chatbot implementations across various educational contexts, aiming to deliver interactive and personalized conversational assistance. For instance, Shaikh et al.~\cite{shaikhchatgptconver23} investigated the use of a ChatGPT-driven chatbot as a conversational support in language learning, revealing its promising effectiveness. In addition, within the gaming industry, ChatGPT-powered Non-Player Characters (NPC) amplify player engagement and enrich gaming experiences~\cite{Huang2023NPCVR}. VR environments stand to benefit significantly from the integration of ChatGPT-powered AI chatbots for conversational support. For example, Nihal et al.~\cite{softskillVRChatgpt23} examined ChatGPT's role as a conversational assistant to promote the development of soft skills in VR, discussing challenges in accuracy, biases, and ethical considerations. Chheang et al.~\cite{genaichat_vr24} conducted a pilot study to examine the effectiveness of a ChatGPT-powered embodied virtual assistant in anatomy education, revealing the significant impact of the chatbot avatar configuration. 
Despite promising prospects, further research is needed to assess the efficacy of integrating ChatGPT-powered chatbots into educational VR applications, understand user perceptions and attitudes, and measure their potential impacts on learning behaviors and outcomes.

\begin{figure*}[t]
\centering
\includegraphics[width=0.90\linewidth]{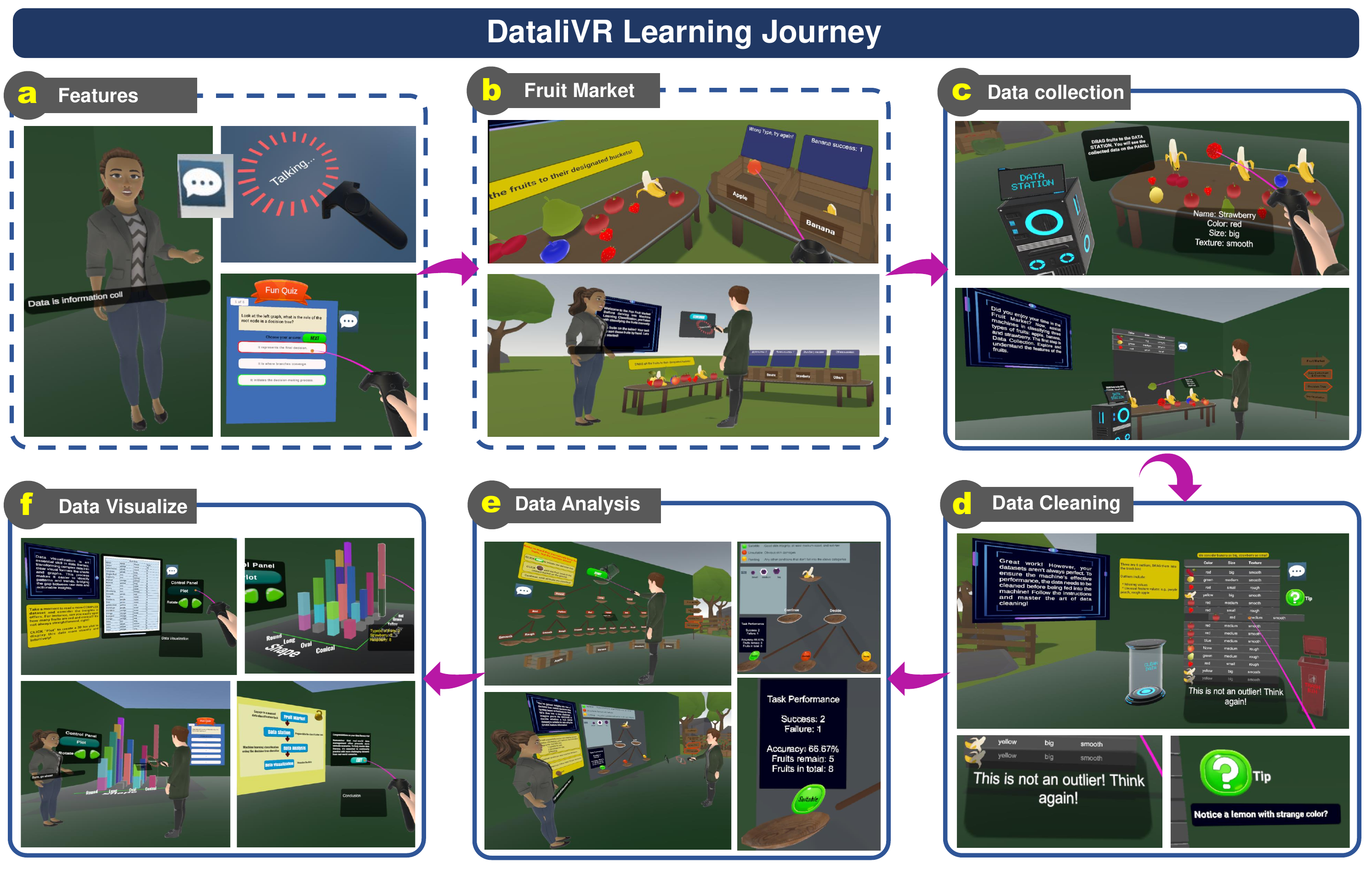} 
\caption{Overview of ChatGPT-enhanced DataliVR}
\label{fig:datalivr}
\end{figure*}

\section{DataliVR Framework}

\subsection{VR Environment Design}
\label{sec:vrdesign}

DataliVR is a comprehensively designed VR application aimed at offering an immersive and unified learning experience in data literacy. It integrates essential aspects of data management, guiding users through key stages including data collection, cleaning, analysis, and visualization, thereby enhancing their data literacy skills. To contextualize data literacy within a tangible framework, we employed a machine learning decision tree classification scenario focused on fruit classification, offering users hands-on engagement throughout the data management process. Tailored for university students with varied backgrounds but not a machine learning background, DataliVR empowers them to develop essential data literacy skills in an engaging environment. The application features five distinct virtual scenes: Fruit Market, Data Collection, Data Cleaning, Data Analysis, and Data Visualization, as illustrated in Figure~\ref{fig:datalivr}. 

\paragraph{Fruit Market: } Before engaging in hands-on data management activities, users are immersed in a simulated Fruit Market environment. Here, they engage in a basic manual fruit classification task. Although seemingly straightforward, this exercise plays an important role in introducing users to basic concepts before moving on to machine learning-based classification. Additionally, quizzes are incorporated to assess users' existing understanding of data literacy and machine learning classification, providing a basis for the subsequent stages of the DataliVR journey. As shown in Figure~\ref{fig:datalivr}b.

\paragraph{Data Collection: }In this scene, users are immersed in a gamified learning environment focused on fruit data collection. A virtual data station is designed to facilitate the interactive process of data collection by allowing users to drag fruits onto the \textit{Data Station}. Once a fruit is selected, an information tag displaying its attributes promptly appears above it. The collected data is then presented in a tabular format on a virtual data panel, reflecting common practices in real data presentation, particularly in machine learning contexts. This immersive experience equips users with hands-on experience in data collection and formatting. As presented in Figure~\ref{fig:datalivr}c.

\paragraph{Data Cleaning: }In this scene, users undergo a comprehensive data-cleaning experience that consists of two phases: passive learning and active exercise. In the learning phase, users engage in guided data-cleaning tasks based on established data-cleaning standards, focusing on cleaning predefined data outliers. Any attempts to remove non-outliers result in the data being returned to the panel along with relevant tips. A help button offers sequential hints about an outlier if none is identified within a set time frame. In the exercise phase, users apply acquired data-cleaning skills to a more complex task, featuring additional fruit attributes. They have the flexibility to remove any data they consider as outliers and can indicate task completion by clicking a finish button. Upon completion, users receive immediate feedback on their task performance and are presented with quizzes to assess their skill acquisition. As shown in Figure~\ref{fig:datalivr}d.

\paragraph{Data Analysis: } In this scene, users delve into data analysis through a machine-learning fruit classification context, specifically centered on a decision tree classifier. Similar to the data cleaning scene, the data analysis session consists of two phases: passive learning and active exercise, aimed at teaching users basic machine-learning classification principles. During the learning phase, users passively interact with a trained decision tree classifier to observe its classification process of various fruit types. They can engage with the classification process by clicking on tree nodes and receiving feedback on performance until all fruits are categorized. In the exercise phase, users engage in a classification task mirroring the decision-making process of a tree classifier to determine whether a fruit is suitable for sale based on the provided decision criteria. Upon completion, users receive immediate feedback on task accuracy, followed by a quiz session. As depicted in Figure~\ref{fig:datalivr}e.

\paragraph{Data Visualization: }In this scene, users explore fruit data using three-dimensional visualizations. They can interact with the 3D plots to extract specific details and compare various aspects of the fruit data. This interactive approach allows users to deepen their understanding of the dataset, making it easier to identify trends and outliers effectively. As users engage with the visualizations, they come across quizzes designed to assess their comprehension of 3D data visualization. As presented in Figure~\ref{fig:datalivr}f.

\subsection{ChatGPT-powered AI Chatbot}
In addition to providing an interactive virtual learning environment enriched with gamified features, DataliVR incorporates a ChatGPT-powered AI Chatbot. For real-time natural language processing and response generation within the virtual setting, we used the ChatGPT API, specifically the GPT-3.5 Turbo version~\cite{gptturbo}. To further enrich user engagement, we integrated Oculus's Voice SDK~\cite{oculusvoicesdk} for text-to-speech conversion and employed OpenAI's Whisper model~\cite{openaiwhisper} for speech-to-text conversion. This integration empowers the chatbot with audio capabilities, enabling a more realistic conversational experience closely mimicking natural dialogue.

As shown in Figure~\ref{fig:datalivr}a, the chatbot was embodied within a virtual avatar. To initiate interaction with the chatbot, users need to activate the avatar by clicking the designed \textit{chat} button. Subsequent interaction is initiated by pressing and holding the controller trigger, releasing it to conclude the conversation. The chatbot's responses also appear in a centrally positioned subtitle panel on the avatar. If no interaction occurs for three seconds, the chatbot avatar fades away. The design of the \textit{chat} button aims to minimize user distraction during the learning process, enabling users to summon the chatbot avatar only when seeking conversational assistance. This design was proven effective through the pilot study.

\section{Method}

\subsection{Participants}
Thirty university students with diverse academic backgrounds and no prior knowledge of machine learning participated in our user study. The participants' fields of study included Educational Science (6), Computer Science (5), Law (3), Engineering (3), Management (3), Math (2), Medicine (2), Psychology (2), Biology (1), Linguistics (1), Environmental Science (1), and Architecture (1). The group consisted of 17 females and 13 males, with an average age of 24.6 ($SD = 2.4$) years. Regarding their prior experience with VR, 33.33\% (10) reported no prior exposure to VR, 36.67\% (11) indicated they had engaged with VR once or twice, 10\% (3) reported three to five VR experiences and 13.33\% (4) reported having more than five VR experiences. This study received approval from the University's Institutional Review Board (IRB). 

\subsection{Apparatus}
We developed DataliVR using the Unity engine (version 2022.3.12f1) in conjunction with SteamVR. Participants experienced the VR environment through the Varjo XR3 headset, which provided a high resolution of 1920 x 1920 pixels per eye and a horizontal field of view of $115^{\circ}$. HTC Vive controllers were employed to facilitate interaction with the VR environment. Varjo base stations were used for headset and controller tracking. The VR system was executed on a high-performance PC equipped with an Intel Core i9 processor, 64GB of RAM, and an NVIDIA GeForce RTX 4080 graphics card, ensuring smooth performance and optimal visual quality throughout the experiment. 

\subsection{Study Design}
Our study aims to comprehensively assess the effectiveness of DataliVR across multiple dimensions of participant learning experiences and outcomes, as detailed in Section~\ref{sec:measures}. In addition, we aim to examine the impact of the ChatGPT-powered AI Chatbot on these aspects. To achieve these objectives, we conducted a controlled experiment using a between-subjects design with a single independent variable, namely the presence of the chatbot, presented at two levels: \textit{\textbf{Without Chatbot}} and \textit{\textbf{With Chatbot}}. Participants were randomly assigned to one of the conditions. 

\subsection{Gamified Tasks}
DataliVR seamlessly combines educational content with entertainment, immersing participants in interactive learning settings centered on data literacy. Participants navigate through a structured sequence of interconnected VR scenes, each presenting unique tasks and challenges related to data collection, cleaning, analysis, and visualization, respectively (see Figure~\ref{fig:datalivr}). As detailed in Section~\ref{sec:vrdesign}, it incorporates passive learning, active exercises, and quiz sessions. Each task is thoughtfully designed, requiring participants to familiarize themselves with the game rules presented in VR and interact with 3D objects using a controller. The sequential progression of scenes captures the excitement of advancing through gaming levels, ensuring an engaging and rewarding experience for participants. Throughout the learning journey, eight exercises were integrated into both the data cleaning and data analysis scenes, accompanied by a total of fifteen quizzes across four scenes. 

\subsection{Measures}
\label{sec:measures}

To evaluate the overall effectiveness of DataliVR, as well as the impact of the ChatGPT-powered AI Chatbot on participants, we employed a comprehensive evaluation approach incorporating both subjective and objective metrics. Our focus centered on two crucial dimensions of participant engagement: learning experiences, and efficiency and outcome. We began with an in-VR pre-test to assess participants' baseline knowledge before they started the virtual learning process. Subsequently, we tracked their learning progression through various tasks integrated within the VR learning modules. After each learning segment, an in-VR quiz session was administered to sustain immersion, eliminating the need for participants to exit the virtual environment. Further details follow below.  

\subsubsection{User Experience}

We evaluated user experience using the User Experience Questionnaire Short Version (UEQs)~\cite{UEQs2017a}. This questionnaire requires a short completion time and includes two scales: one to measure pragmatic quality (task-related) and the other to measure hedonic quality (non-task-related). Each scale consists of four 7-point Likert items, resulting in a total of eight items. 

\subsubsection{System Usability}
We used the System Usability Scale (SUS) questionnaire~\cite{brooke1996sus, brooke2013sus}, a well-established tool for assessing system usability, to assess DataliVR. The SUS comprises ten items that capture key usability aspects, including ease of use, usefulness, perceived complexity, consistency, and ease of learning, each presented as concise statements about the system. Participants indicated their level of agreement with each statement using a 5-point Likert scale. 

\subsubsection{Motion Sickness}

Motion sickness continues to be a significant issue in the VR domain. Therefore, evaluating the degree of motion sickness in users is a crucial aspect of evaluating VR applications' performance, usability, and efficacy. We used a modified version of the Pensacola Diagnostic Criteria survey, known as the Motion Sickness Questionnaire (MSQ)~\cite{motionsickness}. The MSQ contains six items addressing various symptoms, including nausea, cold sweating, drowsiness, headache, flushing, and dizziness. Participants rated their experiences using a 4-point Likert scale, where 1 indicates no symptoms, and 4 indicates severe symptoms. 

\subsubsection{Feedback on Chatbot Experience}

We designed two tailored questionnaires to gather participants' feedback on their experiences with the chatbot and to understand their perceptions in its absence. Each questionnaire comprises six questions, with participants responding using a 5-point Likert scale. For the \textit{Without Chatbot} condition, participants were asked about: 1. the need for additional assistance during learning; 2. the desire for clarification and guidance; 3. seeking support from experimenters; 4. whether the absence of the chatbot hindered learning progress; 5. the impact of the chatbot's absence on their satisfaction; 6. how beneficial they believe a chatbot would have been for assistance during learning. For the \textit{With Chatbot} condition, questions covered: 1. the engagement frequency with the chatbot; 2. enjoyment level while interacting with the chatbot; 3. the chatbot's effectiveness in supporting learning; 4. ease of interaction with the chatbot; 5. the perceived authenticity of the chatbot visualized as an avatar; 6. satisfaction with the chatbot's avatar appearance in VR.  

\subsubsection{Task Completion Time}
The assessment of learning efficiency starts with examining participants' total task completion time. This metric provides a straightforward measure of DataliVR's effectiveness and user-friendliness in improving data literacy skills through descriptive statistics. It evaluates both the speed and proficiency with which participants progress through all phases of the data literacy journey. In this study, the task completion time is defined as the duration participants spend immersed in DataliVR to complete the tasks. 

\subsubsection{Learning Performance}
We evaluated participants' task performance within DataliVR, focusing on their accuracy in completing specific learning tasks, including data cleaning and data analysis in a machine-learning fruit classification context. Moreover, we analyzed their performance in quizzes administered at the end of each data literacy phase to measure their knowledge acquisition. Importantly, participants were unable to directly seek answers to the tasks and quizzes from the chatbot, as these assessments necessitated contextual understanding beyond its designated capabilities. Nonetheless, they could ask for relevant knowledge from the chatbot. 

\subsection{Procedure}
Upon arrival at our VR lab, participants received an introduction to the experiment and then signed a consent form. They were informed about the possibility of experiencing motion sickness during VR immersion and were assured that they could terminate the VR immersion at any time without explanation. Following this, participants completed a demographic questionnaire detailing their age, gender, educational background, and prior VR experience. A pre-test was administered to assess their baseline knowledge of data literacy. Next, participants engaged in a practice session within a separate virtual environment to familiarize themselves with basic VR interactions using the controller and practiced engaging in dialogue with the chatbot. 
The main experiment involved an uninterrupted VR experience in DataliVR. Throughout this immersive session, participants received guided assistance in VR to navigate and complete various learning tasks and exercises, as outlined in Section~\ref{sec:vrdesign}. The duration of this learning session varied among participants, with an average duration of $23.23$ minutes. Upon completion, participants filled out questionnaires about their experiences with DataliVR.
Notably, participants were situated in a solitary lab environment throughout the VR immersion to ease any concerns about feeling monitored, particularly in the chatbot condition, so that they could talk freely with the chatbot and feel comfortable.  

\section{Results}
In this section, we present the results regarding the overall effectiveness of DataliVR and the impact of the chatbot on participants. Our analysis includes descriptive statistics and statistical comparisons between the two conditions using t-tests and Mann-Whitney U tests for parametric and nonparametric tests, respectively. The significance level was set at $\alpha = 0.05$. 

For the questionnaires assessing various user experiences, we calculated each participant's overall score by averaging scores across all items within each questionnaire. For analysis purposes, we reversed scores for certain subscales in some questionnaires, establishing a consistent direction where higher scores consistently indicated either positive or negative experiences. For the learning performance assessment, we reported participants' overall accuracy in task performance across all sixteen exercises and fifteen quizzes within DataliVR. 
Regarding participants' prior knowledge about data literacy, particularly in machine learning classification, we observed lower prior knowledge among participants, with no significant difference between the two conditions ($p>.05$). Additionally, participants' academic backgrounds did not significantly affect their learning experiences and performance ($p>.05$).


\begin{figure*}[t]
  \centering
  \begin{subfigure}{0.3\textwidth} 
    \centering
    \includegraphics[width=\linewidth]{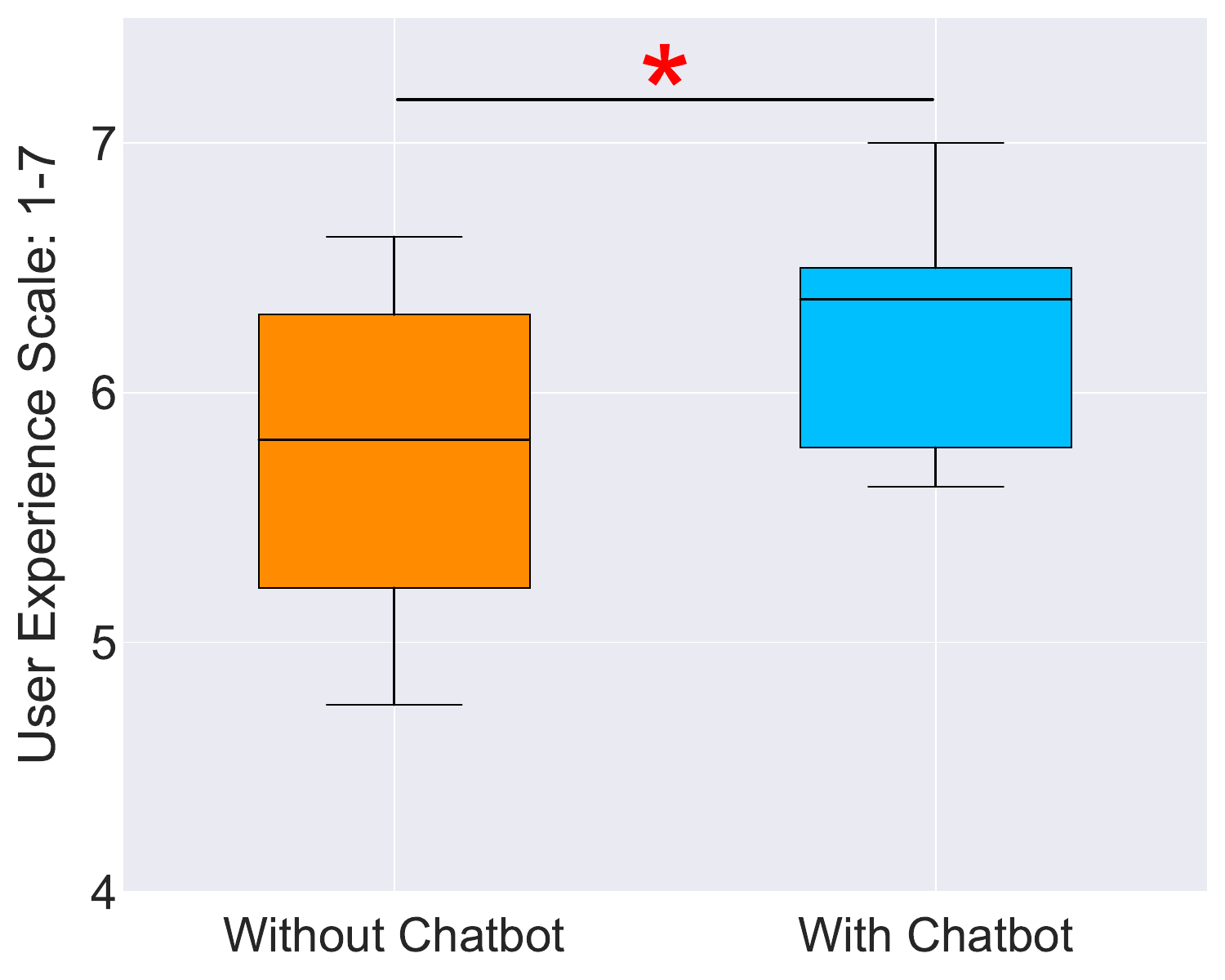} 
    \caption{User experience}
    \label{fig:ue}
  \end{subfigure}
  \begin{subfigure}{0.3\textwidth}
    \centering
    \includegraphics[width=\linewidth]{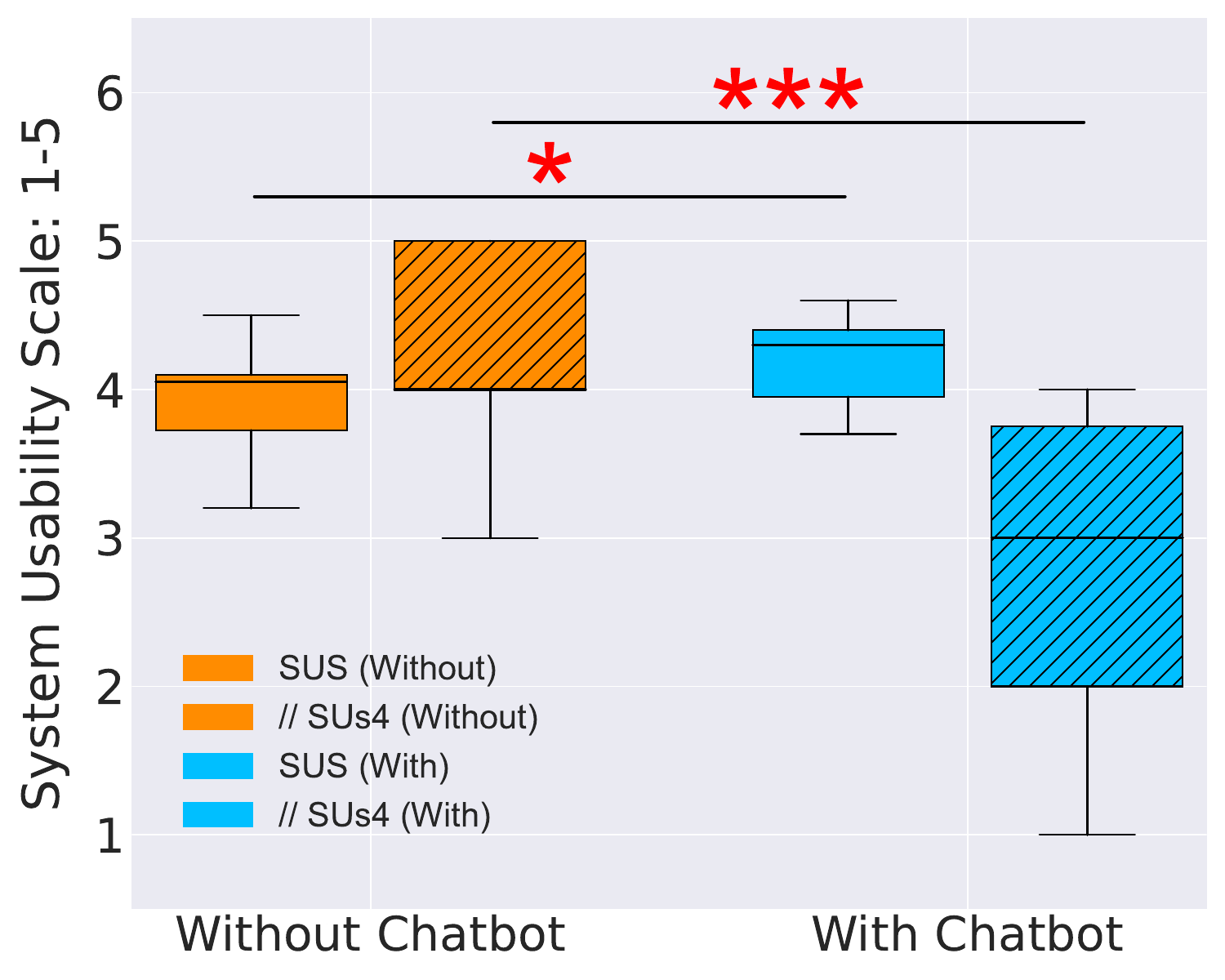} 
    \caption{System usability}
    \label{fig:su}
  \end{subfigure}
  \begin{subfigure}{0.3\textwidth}
    \centering
    \includegraphics[width=\linewidth]{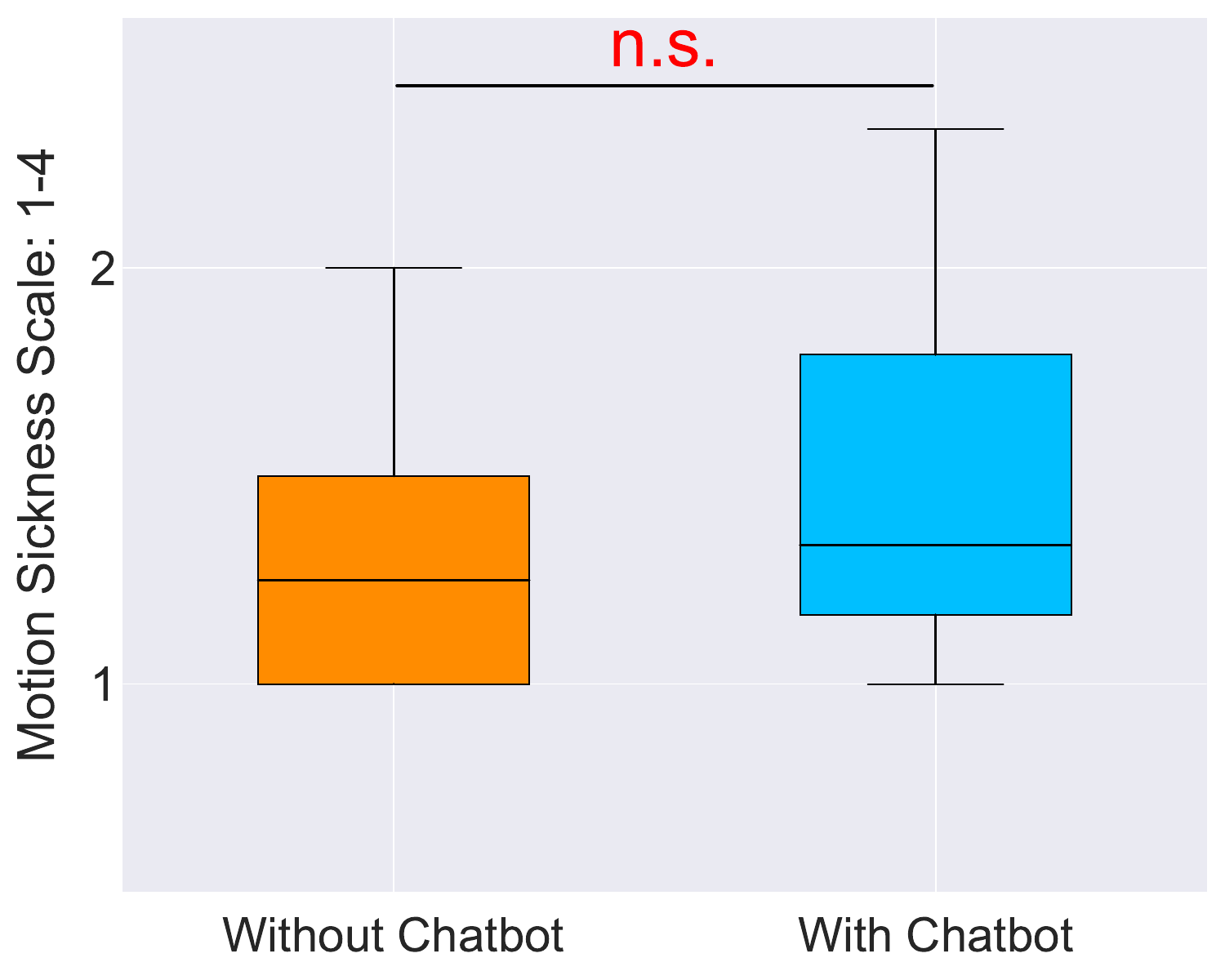} 
    \caption{Motion sickness}
    \label{fig:ms}
  \end{subfigure}
  \caption{User learning experiences in DataliVR}
  \label{fig:overall_ue}
\end{figure*}

\begin{figure*}[!ht]
  \centering
  \begin{subfigure}{0.37\textwidth}
    \centering
    \includegraphics[width=\linewidth]{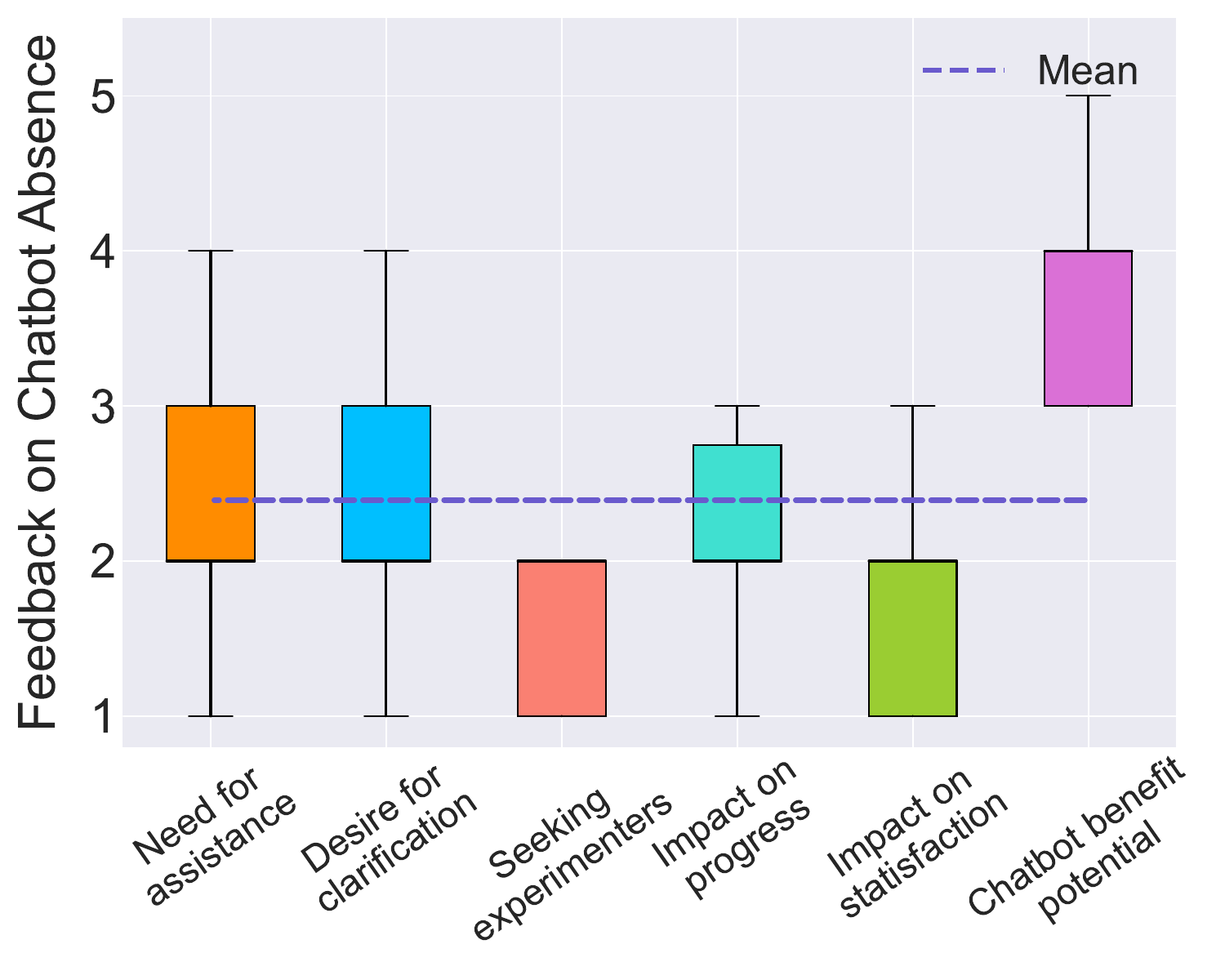} 
    \caption{Feedback on chatbot absence (\textit{Without Chatbot})}
    \label{fig:qc_no}
  \end{subfigure}
  \begin{subfigure}{0.37\textwidth} 
    \centering
    \includegraphics[width=\linewidth]{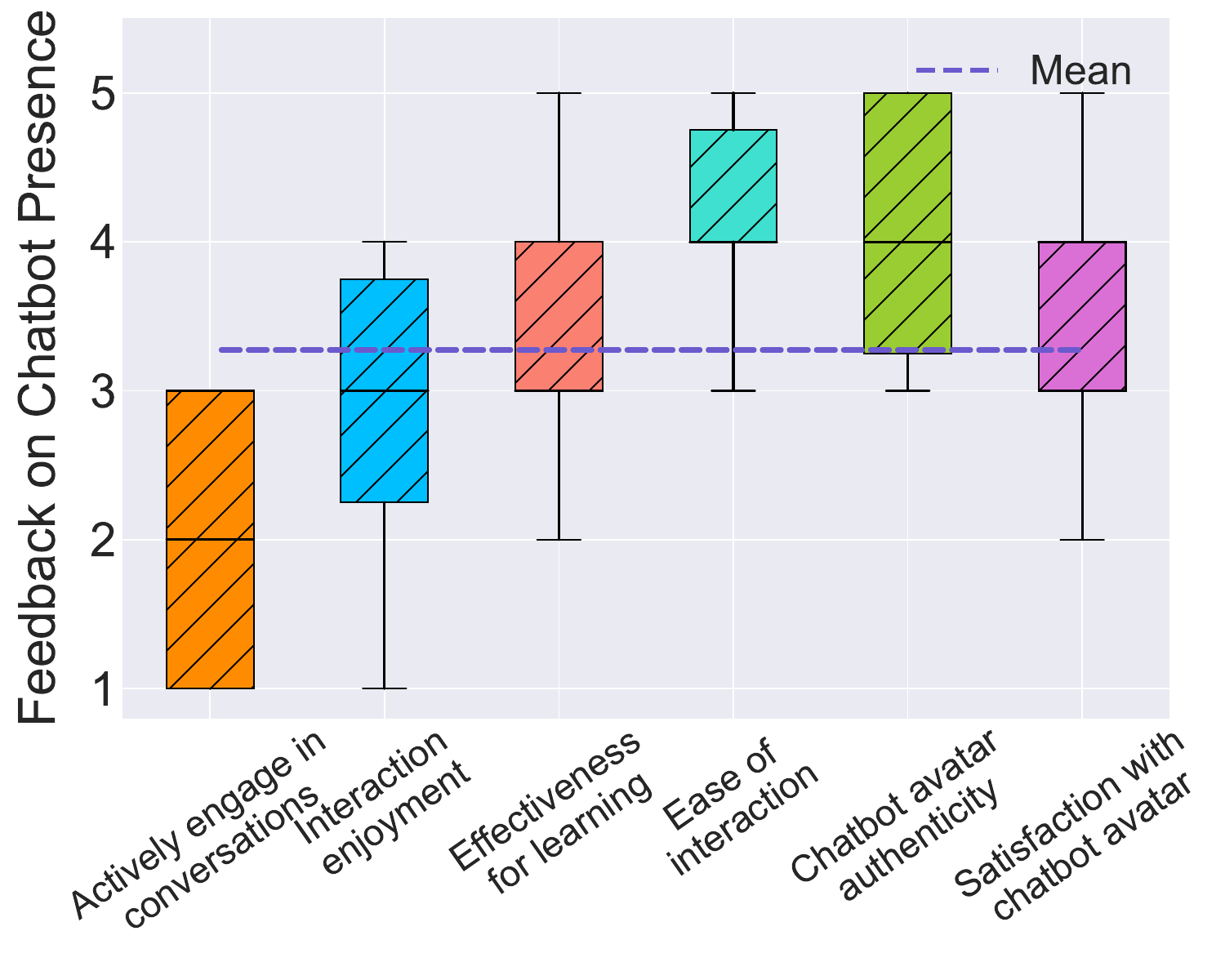} 
    \caption{Feedback on chatbot presence (\textit{With Chatbot})}
    \label{fig:qc_yes}
  \end{subfigure}
  \caption{User feedback on the AI Chatbot}
  \label{fig:overall_qcs}
\end{figure*}

\subsection{User Experience}
The results indicated that participants generally reported a high user experience while using DataliVR, with an average rating of $M=5.99$ ($SD=0.58$) out of 7. Moreover, as shown in Figure~\ref{fig:ue}, we found a significant effect of the chatbot on user experience. Specifically, participants reported a higher level of user experience in the \textit{With Chatbot} condition ($M=6.24$, $SD=0.43$) compared to the \textit{Without Chatbot} condition ($M=5.75$, $SD=0.62$), with $t(28)=-2.3$, $p =.028$. 

\subsection{System Usability}
Similarly, participants generally rated the system usability of DataliVR favorably, indicating its high effectiveness with an average level of $M=4.05$ ($SD=0.36$) out of 5. As depicted in Figure~\ref{fig:su}, the \textit{With Chatbot} condition ($M=4.19$, $SD=0.27$) demonstrated superior usability compared to the \textit{Without Chatbot} condition ($M=3.91$, $SD=0.38$). This difference was statistically significant with $t(28)=-2.12$, $p =.044$, indicating the enhanced user-friendliness facilitated by the presence of the chatbot. 

Interestingly, while participants reported better overall system usability for DataliVR, there was a notable deviation in the SUS-4 item (see Figure~\ref{fig:su}) regarding the need for technical support. Despite the higher SUS scores in the \textit{With Chatbot} condition, participants reported less confidence ($M=2.78$, $SD=0.93$) in using DataliVR without the support of a technical person compared to the \textit{Without Chatbot} condition ($M=4.07$, $SD=1.03$). This difference was statistically significant, as indicated by a Mann-Whitney U statistic of $U=159$, $p =.004$ for a nonparametric comparison.  

\subsection{Motion Sickness}
Participants reported experiencing low motion sickness in DataliVR, averaging a rating of $M=1.42$ ($SD=0.45$) out of 4. Moreover, motion sickness ratings were similar across conditions, with the \textit{Without Chatbot} condition at $M=1.32$ ($SD=0.32$) and the \textit{With Chatbot} condition at $M=1.53$ ($SD=0.53$). No statistically significant differences were observed ($p>.05$) (see Figure~\ref{fig:ms}). 

\begin{figure*}[ht]
  \centering
  \begin{subfigure}{0.3\textwidth} 
    \centering
    \includegraphics[width=\linewidth]{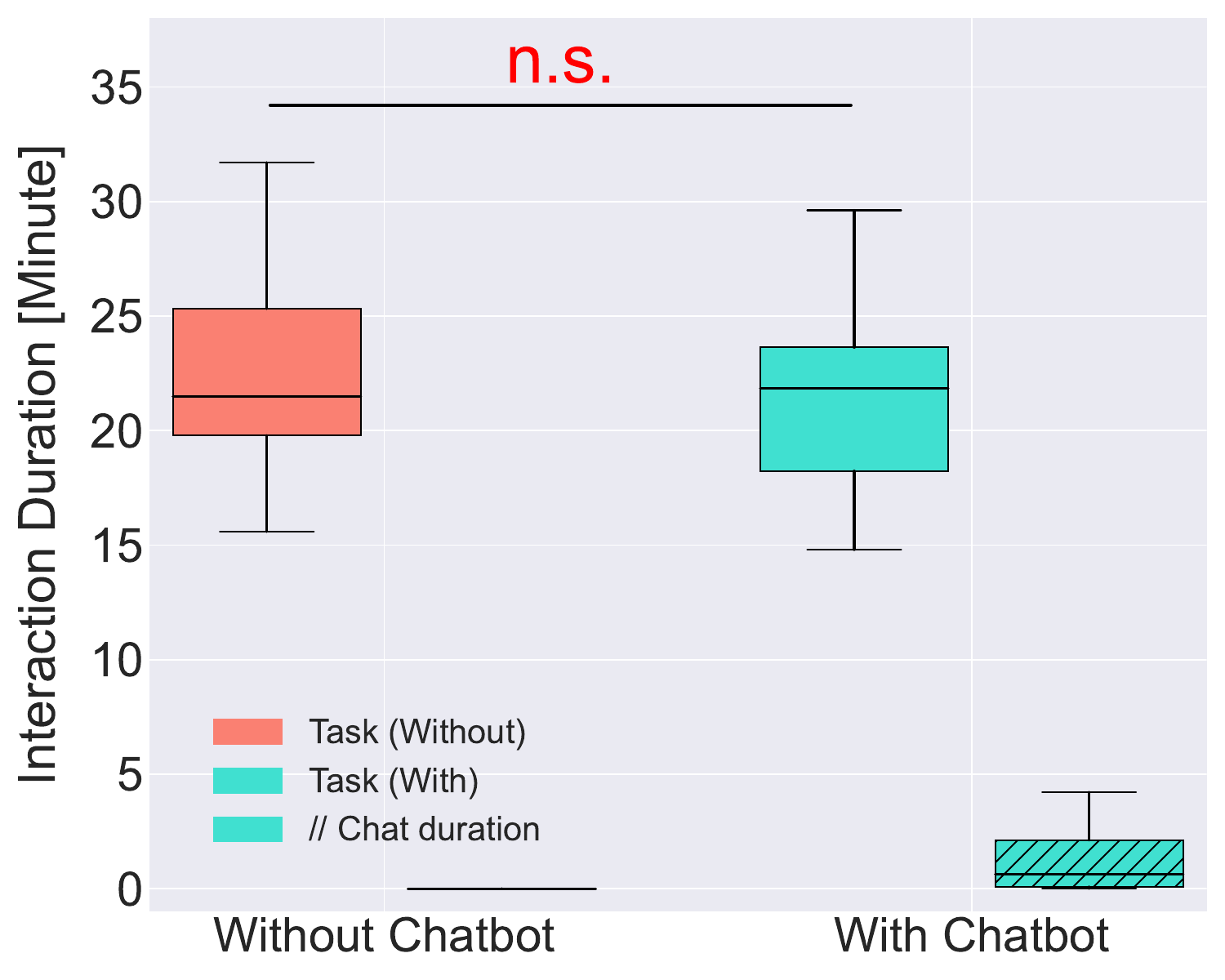} 
    \caption{Task completion time and chat interaction}
    \label{fig:tasktime}
  \end{subfigure}
  \begin{subfigure}{0.3\textwidth} %
    \centering
    \includegraphics[width=\linewidth]{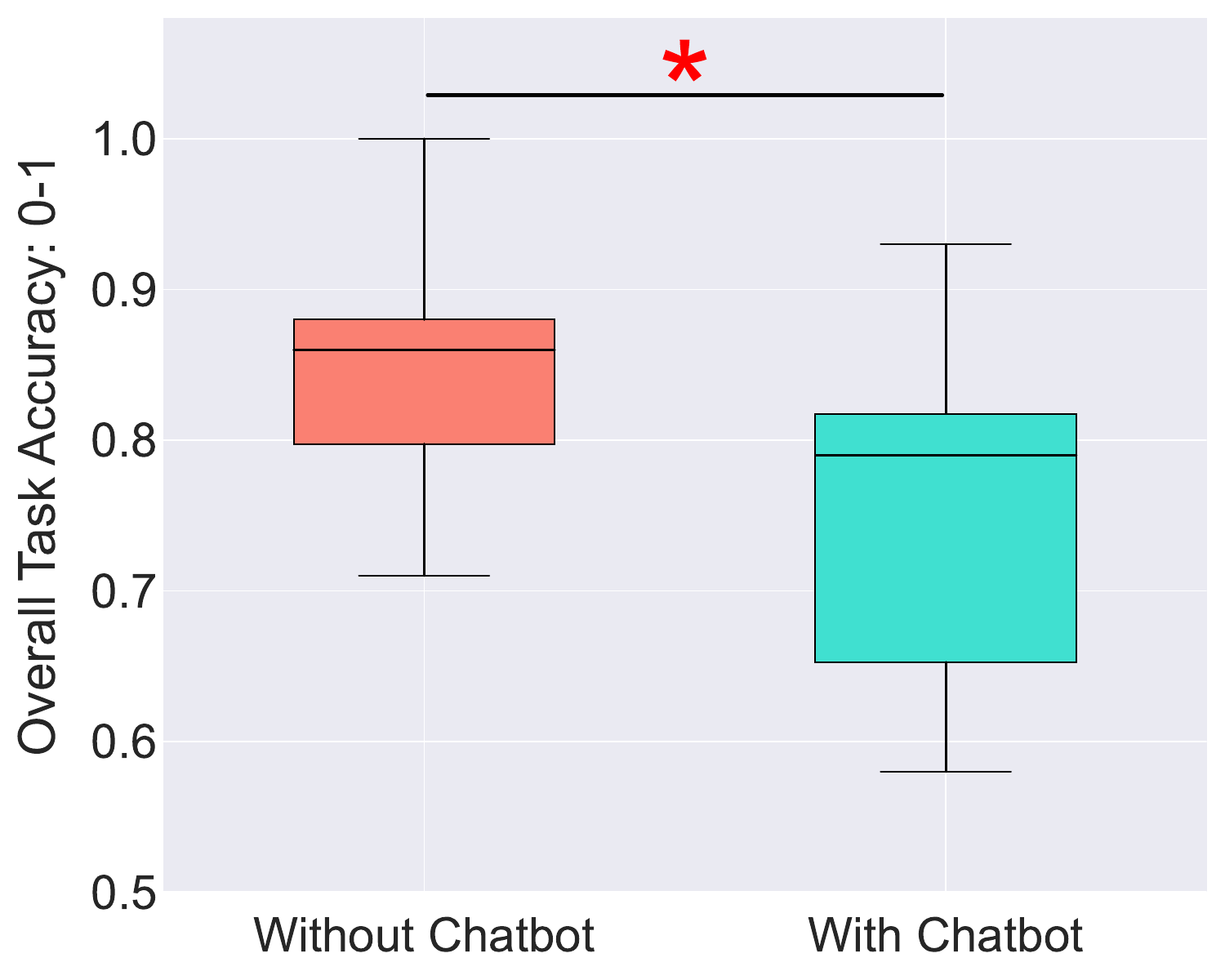} 
    \caption{Overall interactive task accuracy}
    \label{fig:acc_exercise}
  \end{subfigure}
  \begin{subfigure}{0.3\textwidth}
    \centering
    \includegraphics[width=\linewidth]{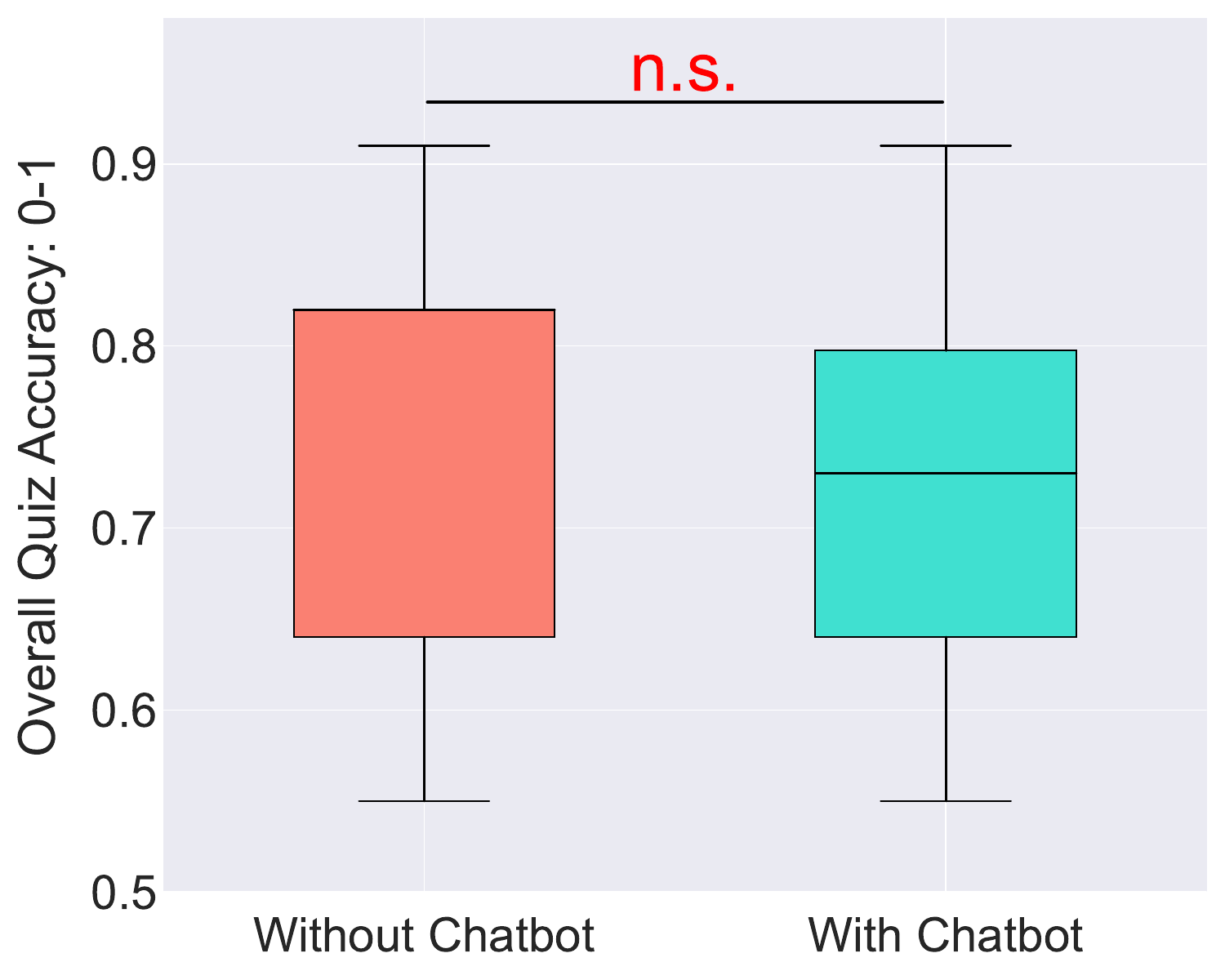} 
    \caption{Overall interactive quiz accuracy}
    \label{fig:acc_quiz}
  \end{subfigure}
  \caption{User learning performance in DataliVR}
  \label{fig:overall_per}
\end{figure*}

\subsection{Feedback on Chatbot Experience}
\label{sec:result_feedbackChatbot}

In the \textit{Without Chatbot} condition, participants provided feedback on the absence of the chatbot across six dimensions. As depicted in Figure~\ref{fig:qc_no}, participants rated the impact of the chatbot's absence on their learning experience as minimal, reflected in items 1 to 5, with an average score of $M=2.11$ ($SD= 0.59$) out of 5. Specifically, participants indicated a limited need for additional assistance ($M=2.21$, $SD=0.89$), clarification of tasks ($M=2.36$, $SD=0.93$), and help from experimenters ($M=1.93$, $SD=1.21$). They also perceived the chatbot's absence as having a minor impact on their learning progression and grasp of learning content($M=2.14$, $SD=0.86$), and on their overall satisfaction with DataliVR ($M=1.93$, $SD=1.07$). Despite these findings, participants still recognized the significant potential benefits of having a chatbot to support learning within DataliVR ($M=3.79$, $SD=0.71$). 

In the \textit{With Chatbot} condition, participants spent an average of $1.48$ minutes engaging with the chatbot, which was relatively low considering the average immersion time of $23.23$ minutes in DataliVR. Notably, four participants did not engage with the chatbot at all, a point further discussed in Section~\ref{sec:disuss_2}. 
Moreover, participants provided feedback on their experience with the chatbot across six dimensions, rating each dimension on a scale of 1 to 5. As shown in Figure~\ref{fig:overall_qcs}, participants rated their overall positive experience with the chatbot at $M=3.27$ ($SD= 0.64$) out of 5. However, as depicted in Figure~\ref{fig:qc_yes}, participants engaged less frequently in conversations with the chatbot ($M=1.93$, $SD=0.92$) and reported moderate enjoyment during interactions ($M=2.86$, $SD=1.03$). Despite this, participants considered the chatbot effective for learning ($M=3.29$, $SD=1.33$) and found it easy to interact with ($M=3.93$, $SD=1.07$). Additionally, participants perceived the chatbot embodied in the virtual avatar as highly authentic ($M=4.07$, $SD=0.83$) and reported high satisfaction with its representation in the virtual environment ($M=3.57$, $SD=1.16$).  

\subsection{Task Completion Time}
The overall task completion time served as a metric to evaluate the efficiency of DataliVR. Immersion time was divided into engagement in tasks and interaction with the chatbot. In the \textit{With Chatbot} condition, the time spent interacting with the chatbot was excluded from the task completion time calculation. As a result, participants spent an average of $22.56$ minutes completing tasks in the \textit{Without Chatbot} condition and $22.49$ minutes in the \textit{With Chatbot} condition. As depicted in Figure~\ref{fig:tasktime}, no statistically significant difference was observed in task completion time ($p>.05$).  

\subsection{Learning Performance}

Participants' learning performance was evaluated from two perspectives: task completion and quiz results. On average, participants showed proficiency in both fruit data cleaning and analysis (machine-learning classification) tasks, achieving an overall average accuracy of $M=0.79$ ($SD=0.10$). Additionally, their performance in knowledge quizzes reflected a robust understanding of data literacy, with an overall average accuracy of $M=0.73$ ($SD=0.11$) across all quizzes.  

Furthermore, the presence of the chatbot was found to impact participants' task performance. Surprisingly, Figure~\ref{fig:acc_exercise} reveals that participants achieved higher task accuracy in the \textit{Without Chatbot} condition ($M=0.83$, $SD=0.07$) compared to the \textit{With Chatbot} condition ($M=0.75$, $SD=0.11$). This difference was statistically significant, with $t(28)=2.26$, $p =.032$. A similar trend was observed in participants' performance in the in-VR quiz sessions, as shown in Figure~\ref{fig:acc_quiz}. In this context as well, participants exhibited better quiz performance in the \textit{Without Chatbot} condition ($M=0.75$, $SD=0.12$) compared to the \textit{With Chatbot} condition ($M=0.71$, $SD=0.11$), although this difference was not statistically significant ($p>.05$). 

\section{Discussion}
\label{sec:discuss}

Our findings underscore the robust usability and effectiveness of DataliVR. The impact of the ChatGPT-powered AI Chatbot on participants' learning varied across different aspects. We discuss our findings in detail below. 

\subsection{Effectiveness of DataliVR in Fostering Data Literacy Learning}
\label{sec:disuss_1}

In this section, we discuss the effectiveness of DataliVR from two key perspectives: delivering user-friendly learning experiences and supporting effective knowledge acquisition. Our findings indicate that participants rated their user experience with DataliVR highly, averaging $M=5.99$ on a scale of 1 to 7. They expressed significant interest in using DataliVR and found the gamified learning journey to be highly beneficial for understanding data literacy concepts and developing relevant skills. In addition, participants rated the system usability of DataliVR favorably, averaging $M=4.05$ on a scale of 1 to 5. They found DataliVR well-designed, featuring a seamless guided learning process, diverse functions, and user-friendliness that minimized the need for additional technical support during use. Despite an average immersion duration of $23.23$ minutes, participants reported experiencing minimal motion sickness, averaging $M=1.42$ on a scale of 1 to 4, indicating a seamless and enjoyable learning experience facilitated by DataliVR.  

Furthermore, the effectiveness of DataliVR is evident in participants' knowledge performance. Our findings reveal high levels of knowledge acquisition in two key aspects: task performance and in-VR quiz accuracies. Participants achieved an average accuracy of $M=0.79$ in performing data cleaning and analysis tasks, and $M=0.73$ in answering quizzes after engaging in passive learning and exercise sessions within each scene. Notably, considering participants' lack of prior knowledge in data literacy and machine learning classification and the challenging nature of the exercises, achieving an overall average accuracy of over $0.75$ in task performance signifies significant learning within DataliVR. 
Overall, our results demonstrate DataliVR's high usability and effectiveness in fostering the development of data literacy skills.

\subsection{Differential Impact of ChatGPT on Learning Aspects}
\label{sec:disuss_2}

This section discusses the effects of the ChatGPT-powered AI chatbot on participants' learning experiences and outcomes and their interactions with the chatbot. Our findings suggest that participants in both conditions generally had positive learning experiences (see Figure~\ref{fig:overall_ue}) with DataliVR. Despite their infrequent interactions with the chatbot, we observed significant impacts of its presence on participants' learning experiences and outcomes. Specifically, DataliVR integrated with the Chatbot provided participants with a significantly better user experience ($p=.028$, see Figure~\ref{fig:ue}). However, it is worth noting that although participants rated the overall system usability of the chatbot-powered DataliVR highly ($p=.044$, see Figure~\ref{fig:su}), they expressed a significantly greater need for technical support to use it ($p=.004$). This discrepancy does not contradict the results indicating a higher general user experience in the \textit{With Chatbot} condition, as the need for technical support specifically relates to the usage of the chatbot. Despite undergoing a flexible practice session, some participants still hesitated to engage with the chatbot due to their limited experience with VR and ChatGPT. For instance, P1 mentioned, \textit{``I have problems interacting with the chatbot. It is not as intelligent as I imagined''}. In the recorded VR scene video, P1 was observed struggling multiple times to interact with the chatbot before eventually having a successful conversation. 

Regarding participants' learning performance, the results show high learning outcomes under both conditions (see Figure~\ref{fig:overall_per}). However, the presence and absence of the chatbot appeared to have different effects on learning outcomes. Surprisingly, in the \textit{Without Chatbot} condition, participants exhibited higher learning performance in both task performances ($p =.032$) and in-VR quiz sessions (though $p >.05$). This finding challenges the expected assumption that the chatbot would enhance learning outcomes. Considering previous user experience feedback (see Figure~\ref{fig:overall_ue}), one possible explanation for this unexpected result is the immersive nature of the designed tasks, and the comprehensiveness of instructions and learning progression in a step-by-step manner, which may have enabled participants to achieve high performance even without additional assistance from the chatbot. Additionally, despite the fluently designed interaction with the avatar-based chatbot, participants rated a higher need for technical support while using DataliVR ($p=.004$, see Figure~\ref{fig:su}), which may be attributed to the chatbot according to participants' posthoc feedback. This could potentially account for lower learning performance. Another factor could be the potential distraction posed by the chatbot's presence during participants' immersion in DataliVR. Although the chatbot avatar is only visible when participants initiate a conversation to reduce distractions, the chatbot button remains prominently visible within the participant's field of view throughout all VR scenes, featuring a jump animation every five seconds (see Figure~\ref{fig:datalivr}a). However, a more in-depth exploration of the relationship between learning performance and the chatbot's presence or absence requires a thorough analysis of participants' interaction videos and audio recordings within DataliVR, as well as learning behavior data in VR. Such an analysis could shed light on how participants engage with the chatbot and how this interaction influences their learning outcomes. While this investigation falls beyond the scope of the current study, it represents a crucial next research step. 

Next, we delve deeper into participants' interaction with the chatbot. Surprisingly, participants in the \textit{With Chatbot} condition engaged in conversations with the chatbot for an average duration of $1.48$ minutes, lower than anticipated. This limited interaction was also reflected in participants' experience feedback, with an average rating of $1.93$ on a scale of 1 to 5 regarding their level of active engagement in chatbot conversations (see Figure~\ref{fig:qc_yes}). Particularly, four participants did not engage with the chatbot at all during their learning in DataliVR. The qualitative feedback from these participants reflects their sense of independence and satisfaction with the learning journey in DataliVR, as expressed by P2: \textit{``I felt totally immersed in performing tasks in DatliVR''}, P6: \textit{``I understood all the instructions presented in DataliVR and can perform tasks myself without additional support from the chatbot''}, P25: \textit{``I don't need support from the chatbot; the learning journey is very smoothly designed''}, and P28: \textit{``I don't know what to ask, and I forgot to ask during my learning and task completion''}. 
However, despite the infrequent interactions with the chatbot, participants consistently rated its presence in DataliVR highly overall (see Figure~\ref{fig:qc_yes}), indicating that they considered the chatbot as positively influencing their learning experiences. They perceived the chatbot as effective, easy to interact with, and authentic, as demonstrated in prior research showing higher performance with a ChatGPT assistant embodied within an avatar compared to a non-avatar condition~\cite{genaichat_vr24}. Moreover, participants found the avatar to be satisfying. These findings are consistent with the broader learning experience in DataliVR, where participants reported high engagement, satisfaction, and usability of DataliVR for skill development (see Figure~\ref{fig:ue} and Figure~\ref{fig:su}). 

Moreover, we also collected participants' feedback on the absence of the chatbot to assess its potential impact on their learning experience. Participants were asked to express whether they sensed the absence of the chatbot, whether they believed their learning was influenced by its absence, and if they anticipated having a chatbot for additional support during learning (see Figure~\ref{fig:qc_no}). Our findings revealed that participants perceived minimal effects of the chatbot's absence on their learning, averaging $2.11$ on a scale of 1 to 5. Specifically, participants expressed a limited level of need for assistance and clarification of tasks from a chatbot. Moreover, the chatbot's absence did not significantly affect participants' learning progression and overall satisfaction. These findings align with earlier findings indicating relatively higher learning performance in the \textit{Without} condition compared to the \textit{With} condition (see Figure~\ref{fig:overall_per}). Nonetheless, participants still expressed that having an AI chatbot for assistance during their learning in DataliVR would be beneficial, rating it at an average of $3.79$ on a scale of 1 to 5. 

Overall, our study demonstrates the potential impact of the ChatGPT-powered AI Chatbot in DataliVR, enhancing participants' learning experiences and influencing their learning outcomes. However, further exploration is needed to fully understand the relationship between the chatbot's presence and learning outcomes. Our findings offer valuable insights for future research aiming to integrate ChatGPT-powered AI Chatbots into educational VR applications. Rather than simply using ChatGPT for its capabilities and popularity, it is crucial to delve deeper into users' perceptions and attitudes toward engaging with ChatGPT. Despite ChatGPT's popularity, many individuals, especially those not from computer science, still do not utilize it on a daily basis. Moreover, our findings highlight how the immersive nature of tasks in VR can impact users' interaction with the chatbot. Users may become deeply immersed in tasks, potentially overlooking the chatbot's role. To address this, researchers can implement strategies to promote interaction with ChatGPT. For instance, introducing more knowledge-related challenges rather than enabling users to summon the chatbot avatar only when seeking conversational assistance. This approach would encourage users to actively seek guidance from ChatGPT, thereby enhancing learning outcomes through ChatGPT's contribution to knowledge acquisition and effectiveness in learning. 

\section{Conclusion}
This study introduces DataliVR, an immersive and interactive VR application developed to enhance data literacy, enriched by the integration of a ChatGPT-powered AI Chatbot. DataliVR offers participants a comprehensive data literacy learning experience, covering aspects of data collection, cleaning, analysis, and visualization. Our extensive evaluations of DataliVR reveal its high effectiveness and user-friendliness across various dimensions of user experience and learning outcomes. 
Moreover, our findings reveal the positive impact of the chatbot on users' learning experiences, underscoring its considerable potential for enhancing educational VR applications. However, the chatbot did not yield a significantly positive effect on learning outcomes, necessitating further investigation into their underlying relationships before drawing definitive conclusions. This underscores the importance of researchers exploring the integration of ChatGPT in supporting VR learning to delve deeper into its potential implications. In summary, our study not only contributes a valuable tool for advancing data literacy education and driving the digital transformation of this field but also offers important insights for the future development of ChatGPT-enhanced educational VR applications and beyond.

\bibliographystyle{abbrv-doi}

\bibliography{reference}

\begin{thebibliography}{10}

\bibitem{vrcheconcept21}
M.~Abdinejad, B.~Talaie, H.~S. Qorbani, and S.~Dalili.
\newblock Student perceptions using augmented reality and 3d visualization technologies in chemistry education.
\newblock {\em Journal of Science Education and Technology}, 30:87--96, 2021.

\bibitem{reviewmrteacher22}
G.~O. Ade-Ojo, M.~Markowski, R.~Essex, M.~Stiell, and J.~Jameson.
\newblock A systematic scoping review and textual narrative synthesis of physical and mixed-reality simulation in pre-service teacher training.
\newblock {\em Journal of Computer Assisted Learning}, 38(3):861--874, 2022.

\bibitem{vrmath23}
E.~Akman and R.~Çakır.
\newblock The effect of educational virtual reality game on primary school students’ achievement and engagement in mathematics.
\newblock {\em Interactive Learning Environments}, 31(3):1467--1484, 2023.

\bibitem{ivrproblemsolve21}
P.~Araiza-Alba, T.~Keane, W.~S. Chen, and J.~Kaufman.
\newblock Immersive virtual reality as a tool to learn problem-solving skills.
\newblock {\em Computers \& Education}, 164:104121, 2021.

\bibitem{bas2023language}
R.~Baskara et~al.
\newblock Exploring the implications of chatgpt for language learning in higher education.
\newblock {\em Indonesian Journal of English Language Teaching and Applied Linguistics}, 7(2):343--358, 2023.

\bibitem{brooke1996sus}
J.~Brooke.
\newblock Sus: a “quick and dirty’usability.
\newblock {\em Usability evaluation in industry}, 189(3):189--194, 1996.

\bibitem{brooke2013sus}
J.~Brooke.
\newblock Sus: a retrospective.
\newblock {\em Journal of usability studies}, 8(2):29--40, 2013.

\bibitem{vreduremote22}
C.-S. Chan, J.~Bogdanovic, and V.~Kalivarapu.
\newblock Applying immersive virtual reality for remote teaching architectural history.
\newblock {\em Education and Information Technologies}, 27(3):4365--4397, 2022.

\bibitem{ChenVRteacherpilot22}
C.-Y. Chen.
\newblock Immersive virtual reality to train preservice teachers in managing students' challenging behaviours: A pilot study.
\newblock {\em British Journal of Educational Technology}, 53(4):998--1024, 2022.

\bibitem{GPTutor23Chen}
E.~Chen, R.~Huang, H.-S. Chen, Y.-H. Tseng, and L.-Y. Li.
\newblock Gptutor: A chatgpt-powered programming tool for code explanation.
\newblock In {\em Artificial Intelligence in Education. Posters and Late Breaking Results, Workshops and Tutorials, Industry and Innovation Tracks, Practitioners, Doctoral Consortium and Blue Sky}, pp. 321--327. Springer Nature Switzerland, Cham, 2023. doi: {{%
10\hspace{.1pt}\discretionary{.}{%
}{.}\hspace{.4pt}1007\discretionary{/}{%
}{/}978\discretionary{%
}{-}{-}3\discretionary{%
}{-}{-}031\discretionary{%
}{-}{-}36336\discretionary{%
}{-}{-}8\_50}}


\bibitem{genaichat_vr24}
V.~Chheang, S.~Sharmin, R.~Márquez-Hernández, M.~Patel, D.~Rajasekaran, G.~Caulfield, B.~Kiafar, J.~Li, P.~Kullu, and R.~L. Barmaki.
\newblock Towards anatomy education with generative ai-based virtual assistants in immersive virtual reality environments.
\newblock In {\em 2024 IEEE International Conference on Artificial Intelligence and eXtended and Virtual Reality (AIxVR)}, pp. 21--30, 2024.

\bibitem{dlassessment}
Y.~Cui, F.~Chen, A.~Lutsyk, J.~P. Leighton, and M.~Cutumisu.
\newblock Data literacy assessments: a systematic literature review.
\newblock {\em Assessment in Education: Principles, Policy \& Practice}, 30(1):76--96, 2023.

\bibitem{vrlanguage22review}
T.~K. Dhimolea, R.~Kaplan-Rakowski, and L.~Lin.
\newblock A systematic review of research on high-immersion virtual reality for language learning.
\newblock {\em TechTrends}, 66(5):810--824, 2022.

\bibitem{vrpairprogram20}
J.~Dominic, B.~Tubre, C.~Ritter, J.~Houser, C.~Smith, and P.~Rodeghero.
\newblock Remote pair programming in virtual reality.
\newblock In {\em 2020 IEEE International Conference on Software Maintenance and Evolution (ICSME)}, pp. 406--417, 2020.

\bibitem{datalidef2016}
K.~Dunlap and J.~S. Piro.
\newblock Diving into data: Developing the capacity for data literacy in teacher education.
\newblock {\em Cogent Education}, 3(1):1132526, 2016.

\bibitem{dlhighereducation}
J.~Elisa~Raffaghelli.
\newblock Is data literacy a catalyst of social justice? a response from nine data literacy initiatives in higher education.
\newblock {\em Education Sciences}, 10(9), 2020.

\bibitem{Gao_VRclass_chi21}
H.~Gao, E.~Bozkir, L.~Hasenbein, J.-U. Hahn, R.~G\"{o}llner, and E.~Kasneci.
\newblock Digital transformations of classrooms in virtual reality.
\newblock In {\em Proceedings of the 2021 CHI Conference on Human Factors in Computing Systems}, CHI '21. ACM, New York, NY, USA, 2021.

\bibitem{Gao_VRteacher23}
H.~Gao, E.~Bozkir, P.~Stark, P.~Goldberg, G.~Meixner, E.~Kasneci, and R.~Göllner.
\newblock Detecting teacher expertise in an immersive vr classroom: Leveraging fused sensor data with explainable machine learning models.
\newblock In {\em 2023 IEEE International Symposium on Mixed and Augmented Reality (ISMAR)}, pp. 683--692, 2023.

\bibitem{gao2023VRgender}
H.~Gao, L.~Hasenbein, E.~Bozkir, R.~G{\"o}llner, and E.~Kasneci.
\newblock Exploring gender differences in computational thinking learning in a vr classroom: Developing machine learning models using eye-tracking data and explaining the models.
\newblock {\em International Journal of Artificial Intelligence in Education}, 33(4):929--954, 2023.

\bibitem{vrphysic21}
Y.~Georgiou, O.~Tsivitanidou, and A.~Ioannou.
\newblock Learning experience design with immersive virtual reality in physics education.
\newblock {\em Educational Technology Research and Development}, 69(6):3051--3080, 2021.

\bibitem{datamange18}
A.~Grillenberger and R.~Romeike.
\newblock Developing a theoretically founded data literacy competency model.
\newblock In {\em Proceedings of the 13th Workshop in Primary and Secondary Computing Education}, WiPSCE '18. ACM, New York, NY, USA, 2018. doi: {{%
10\hspace{.1pt}\discretionary{.}{%
}{.}\hspace{.4pt}1145\discretionary{/}{%
}{/}3265757\hspace{.1pt}\discretionary{.}{%
}{.}\hspace{.4pt}3265766}}


\bibitem{llms23survey}
M.~U. Hadi, R.~Qureshi, A.~Shah, M.~Irfan, A.~Zafar, M.~B. Shaikh, N.~Akhtar, J.~Wu, S.~Mirjalili, et~al.
\newblock A survey on large language models: Applications, challenges, limitations, and practical usage.
\newblock {\em Authorea Preprints}, 2023.

\bibitem{hasenbein22VRsocial}
L.~Hasenbein, P.~Stark, U.~Trautwein, A.~C.~M. Queiroz, J.~Bailenson, J.-U. Hahn, and R.~Göllner.
\newblock Learning with simulated virtual classmates: Effects of social-related configurations on students’ visual attention and learning experiences in an immersive virtual reality classroom.
\newblock {\em Computers in Human Behavior}, 133:107282, 2022.

\bibitem{vravatarchildren22}
S.~Z. Hassan, P.~Salehi, R.~K. R\o{}ed, P.~Halvorsen, G.~A. Baugerud, M.~S. Johnson, P.~Lison, M.~Riegler, M.~E. Lamb, C.~Griwodz, and S.~S. Sabet.
\newblock Towards an ai-driven talking avatar in virtual reality for investigative interviews of children.
\newblock In {\em Proceedings of the 2nd Workshop on Games Systems}, GameSys '22, p. 9–15. ACM, New York, NY, USA, 2022.

\bibitem{dledupro21}
J.~Henderson and M.~Corry.
\newblock Data literacy training and use for educational professionals.
\newblock {\em Journal of Research in Innovative Teaching \& Learning}, 14(2):232--244, 2021.

\bibitem{hu2021VRchemi}
E.~Hu-Au and S.~Okita.
\newblock Exploring differences in student learning and behavior between real-life and virtual reality chemistry laboratories.
\newblock {\em Journal of Science Education and Technology}, 30(6):862--876, 2021.

\bibitem{Huang2023NPCVR}
J.~Huang and K.~Huang.
\newblock {\em ChatGPT in Gaming Industry}, pp. 243--269.
\newblock Springer Nature Switzerland, Cham, 2023.

\bibitem{vrcriticalthink20}
J.~Ikhsan, K.~Sugiyarto, and T.~Astuti.
\newblock Fostering student’s critical thinking through a virtual reality laboratory.
\newblock 2020.

\bibitem{suGPTschoolcontent23}
J.~S. Jauhiainen and A.~G. Guerra.
\newblock Generative ai and chatgpt in school children’s education: Evidence from a school lesson.
\newblock {\em Sustainability}, 15(18), 2023.

\bibitem{vrmedical20}
M.~Javaid and A.~Haleem.
\newblock Virtual reality applications toward medical field.
\newblock {\em Clinical Epidemiology and Global Health}, 8(2):600--605, 2020.

\bibitem{KASNECI2023102274}
E.~Kasneci, K.~Sessler, S.~Küchemann, M.~Bannert, D.~Dementieva, F.~Fischer, U.~Gasser, G.~Groh, S.~Günnemann, E.~Hüllermeier, S.~Krusche, G.~Kutyniok, T.~Michaeli, C.~Nerdel, J.~Pfeffer, O.~Poquet, M.~Sailer, A.~Schmidt, T.~Seidel, M.~Stadler, J.~Weller, J.~Kuhn, and G.~Kasneci.
\newblock Chatgpt for good? on opportunities and challenges of large language models for education.
\newblock {\em Learning and Individual Differences}, 103:102274, 2023.

\bibitem{llmscode24}
M.~Kazemitabaar, X.~Hou, A.~Henley, B.~J. Ericson, D.~Weintrop, and T.~Grossman.
\newblock How novices use llm-based code generators to solve cs1 coding tasks in a self-paced learning environment.
\newblock In {\em Proceedings of the 23rd Koli Calling International Conference on Computing Education Research}, Koli Calling '23. ACM, New York, NY, USA, 2024.

\bibitem{vrart22}
Y.~Kim and H.~Lee.
\newblock Falling in love with virtual reality art: A new perspective on 3d immersive virtual reality for future sustaining art consumption.
\newblock {\em International Journal of Human–Computer Interaction}, 38(4):371--382, 2022.

\bibitem{dl_reandli17}
T.~Koltay.
\newblock Data literacy for researchers and data librarians.
\newblock {\em Journal of Librarianship and Information Science}, 49(1):3--14, 2017.

\bibitem{chatgptTeacher23}
S.~K\"uchemann, S.~Steinert, N.~Revenga, M.~Schweinberger, Y.~Dinc, K.~E. Avila, and J.~Kuhn.
\newblock Can chatgpt support prospective teachers in physics task development?
\newblock {\em Phys. Rev. Phys. Educ. Res.}, 19:020128, Sep 2023.

\bibitem{motionsickness}
B.~D. Lawson, D.~A. Graeber, A.~M. Mead, and E.~R. Muth.
\newblock Signs and symptoms of human syndromes associated with synthetic experiences.
\newblock In {\em Handbook of virtual environments}, pp. 629--658. CRC Press, 2002.

\bibitem{vrscience19}
G.~Makransky, T.~S. Terkildsen, and R.~E. Mayer.
\newblock Adding immersive virtual reality to a science lab simulation causes more presence but less learning.
\newblock {\em Learning and Instruction}, 60:225--236, 2019.

\bibitem{vreduper24}
A.~Marougkas, C.~Troussas, A.~Krouska, and C.~Sgouropoulou.
\newblock How personalized and effective is immersive virtual reality in education? a systematic literature review for the last decade.
\newblock {\em Multimedia Tools and Applications}, 83(6):18185--18233, 2024.

\bibitem{reviewVRStu14}
Z.~Merchant, E.~T. Goetz, L.~Cifuentes, W.~Keeney-Kennicutt, and T.~J. Davis.
\newblock Effectiveness of virtual reality-based instruction on students' learning outcomes in k-12 and higher education: A meta-analysis.
\newblock {\em Computers \& Education}, 70:29--40, 2014.

\bibitem{k12aiml23}
T.~Michaeli, R.~Romeike, and S.~Seegerer.
\newblock What students can learn about artificial intelligence -- recommendations for k-12 computing education.
\newblock In T.~Keane, C.~Lewin, T.~Brinda, and R.~Bottino, eds., {\em Towards a Collaborative Society Through Creative Learning}, pp. 196--208. Springer Nature Switzerland, Cham, 2023.

\bibitem{chatgptessay23}
A.~Mizumoto and M.~Eguchi.
\newblock Exploring the potential of using an ai language model for automated essay scoring.
\newblock {\em Research Methods in Applied Linguistics}, 2(2):100050, 2023.

\bibitem{survey23chatgptagent}
A.~Nazir and Z.~Wang.
\newblock A comprehensive survey of chatgpt: Advancements, applications, prospects, and challenges.
\newblock {\em Meta-Radiology}, 1(2):100022, 2023.

\bibitem{softskillVRChatgpt23}
K.~S. Nihal, L.~Pallavi, R.~Raj, C.~M. Babu, and B.~Mishra.
\newblock Enhancing soft skill development with chatgpt and vr: An exploratory study.
\newblock In {\em 2023 International Conference on Research Methodologies in Knowledge Management, Artificial Intelligence and Telecommunication Engineering (RMKMATE)}, pp. 1--6, 2023.

\bibitem{oculusvoicesdk}
{Oculus}.
\newblock {Oculus Voice SDK}.
\newblock https://developer.oculus.com, 2024.

\bibitem{oguguo2020assessment}
B.~Oguguo, F.~A. Nannim, A.~O. Okeke, R.~I. Ezechukwu, G.~A. Christopher, and C.~O. Ugorji.
\newblock Assessment of students’ data literacy skills in southern nigerian universities.
\newblock {\em Universal Journal of Educational Research}, 8(6):2717--2726, 2020.

\bibitem{chatgpt}
{OpenAI}.
\newblock {ChatGPT}: A large-scale transformer-based conversational model.
\newblock \url{https://openai.com/chatgpt}, 2022.
\newblock Accessed: 2024-05-01.

\bibitem{openaiwhisper}
{OpenAI}.
\newblock {Whisper: Speech Recognition Model}.
\newblock https://platform.openai.com/docs/models/whisper, 2022.

\bibitem{gptturbo}
{OpenAI}.
\newblock {GPT-3.5-turbo}.
\newblock https://platform.openai.com/docs/models/gpt-3-5-turbo, 2023.

\bibitem{vredugame20}
S.~S. Oyelere, N.~Bouali, R.~Kaliisa, G.~Obaido, A.~A. Yunusa, and E.~R. Jimoh.
\newblock Exploring the trends of educational virtual reality games: a systematic review of empirical studies.
\newblock {\em Smart Learning Environments}, 7:1--22, 2020.

\bibitem{vrstcsedu20}
J.~Pirker, A.~Dengel, M.~Holly, and S.~Safikhani.
\newblock Virtual reality in computer science education: A systematic review.
\newblock In {\em Proceedings of the 26th ACM Symposium on Virtual Reality Software and Technology}, VRST'20. ACM, New York, NY, USA, 2020.

\bibitem{dlprogram13}
J.~C. Prado and M.~Ángel Marzal.
\newblock Incorporating data literacy into information literacy programs: Core competencies and contents.
\newblock {\em Libri}, 63(2):123--134, 2013.

\bibitem{rVRedu2020}
J.~Radianti, T.~A. Majchrzak, J.~Fromm, and I.~Wohlgenannt.
\newblock A systematic review of immersive virtual reality applications for higher education: Design elements, lessons learned, and research agenda.
\newblock {\em Computers \& Education}, 147:103778, 2020.

\bibitem{vrhistoryRoman22}
J.~A. G.-C. Rafael Villena~Taranilla, Ramón Cózar-Gutiérrez and I.~L. Cirugeda.
\newblock Strolling through a city of the roman empire: an analysis of the potential of virtual reality to teach history in primary education.
\newblock {\em Interactive Learning Environments}, 30(4):608--618, 2022.

\bibitem{dleduhigher}
J.~E. Raffaghelli and B.~Stewart.
\newblock Centering complexity in ‘educators’ data literacy’ to support future practices in faculty development: a systematic review of the literature.
\newblock {\em Teaching in Higher Education}, 25(4):435--455, 2020.

\bibitem{reevesdlassess19}
T.~D. Reeves and J.-L. Chiang.
\newblock Effects of an asynchronous online data literacy intervention on pre-service and in-service educators’ beliefs, self-efficacy, and practices.
\newblock {\em Computers \& Education}, 136:13--33, 2019.

\bibitem{dldef2015}
C.~Ridsdale, J.~Rothwell, M.~Smit, H.~Ali-Hassan, M.~Bliemel, D.~Irvine, D.~Kelley, S.~Matwin, and B.~Wuetherick.
\newblock Strategies and best practices for data literacy education: Knowledge synthesis report.
\newblock 2015.

\bibitem{UEQs2017a}
M.~Schrepp, J.~Thomaschewski, and A.~Hinderks.
\newblock Design and evaluation of a short version of the user experience questionnaire (ueq-s).
\newblock {\em International Journal of Interactive Multimedia and Artificial Intelligence}, 4(6):103--108, 12/2017 2017.

\bibitem{VRockspro20}
R.~J. Segura, F.~J. del Pino, C.~J. Ogáyar, and A.~J. Rueda.
\newblock Vr-ocks: A virtual reality game for learning the basic concepts of programming.
\newblock {\em Computer Applications in Engineering Education}, 28(1):31--41, 2020.

\bibitem{KathrinChatgptwri23}
K.~Se{\ss}ler, T.~Xiang, L.~Bogenrieder, and E.~Kasneci.
\newblock Peer: Empowering writing with large language models.
\newblock In O.~Viberg, I.~Jivet, P.~J. Mu{\~{n}}oz-Merino, M.~Perifanou, and T.~Papathoma, eds., {\em Responsive and Sustainable Educational Futures}, pp. 755--761. Springer Nature Switzerland, Cham, 2023.

\bibitem{shaikhchatgptconver23}
S.~Shaikh, S.~Y. Yayilgan, B.~Klimova, and M.~Pikhart.
\newblock Assessing the usability of chatgpt for formal english language learning.
\newblock {\em European Journal of Investigation in Health, Psychology and Education}, 13(9):1937--1960, 2023.

\bibitem{vrcomskill20train}
S.~Shorey, E.~Ang, E.~D. Ng, J.~Yap, L.~S.~T. Lau, and C.~K. Chui.
\newblock Communication skills training using virtual reality: A descriptive qualitative study.
\newblock {\em Nurse Education Today}, 94:104592, 2020.

\bibitem{simon22DLcurriculum}
M.~Simon, E.~Prather, I.~Rosenthal, M.~Cassidy, J.~Hammerman, and L.~Trouille.
\newblock A new curriculum development model for improving undergraduate students’ data literacy and self-efficacy in online astronomy classrooms.
\newblock 2022. doi: {{%
10\hspace{.1pt}\discretionary{.}{%
}{.}\hspace{.4pt}32374\discretionary{/}{%
}{/}AEJ\hspace{.1pt}\discretionary{.}{%
}{.}\hspace{.4pt}2022\hspace{.1pt}\discretionary{.}{%
}{.}\hspace{.4pt}2\hspace{.1pt}\discretionary{.}{%
}{.}\hspace{.4pt}1\hspace{.1pt}\discretionary{.}{%
}{.}\hspace{.4pt}043ra}}


\bibitem{vrchemistry23}
R.~van Dinther, L.~de~Putter, and B.~Pepin.
\newblock Features of immersive virtual reality to support meaningful chemistry education.
\newblock {\em Journal of Chemical Education}, 100(4):1537--1546, 2023.

\bibitem{villagptagent23}
L.~Villa, D.~Carneros-Prado, A.~S{\'a}nchez-Miguel, C.~C. Dobrescu, and R.~Herv{\'a}s.
\newblock Conversational agent development through large language models: Approach with gpt.
\newblock In {\em Proceedings of the 15th International Conference on Ubiquitous Computing {\&} Ambient Intelligence (UCAmI 2023)}, pp. 286--297. Springer Nature Switzerland, Cham, 2023.

\bibitem{wolff16DLurban}
A.~Wolff, J.~J.~C. Montaner, and G.~Kortuem.
\newblock Urban data in the primary classroom: bringing data literacy to the uk curriculum.
\newblock {\em The Journal of Community Informatics}, 12(3), 2016.

\bibitem{Wu20VRreview}
B.~Wu, X.~Yu, and X.~Gu.
\newblock Effectiveness of immersive virtual reality using head-mounted displays on learning performance: A meta-analysis.
\newblock {\em British Journal of Educational Technology}, 51(6):1991--2005, 2020.

\bibitem{wu23dataAnateacher}
X.~Wu, T.~Xu, and Y.~Zhang.
\newblock Research on the data analysis knowledge assessment of pre-service teachers from china based on cognitive diagnostic assessment.
\newblock {\em Current Psychology}, 42(6):4885--4899, 2023.

\bibitem{rVR_skilltrain21}
B.~Xie, H.~Liu, R.~Alghofaili, Y.~Zhang, Y.~Jiang, F.~D. Lobo, C.~Li, W.~Li, H.~Huang, M.~Akdere, C.~Mousas, and L.-F. Yu.
\newblock A review on virtual reality skill training applications.
\newblock {\em Frontiers in Virtual Reality}, 2, 2021.

\bibitem{youngChatdialogue23}
J.~C. Young and M.~Shishido.
\newblock Investigating openai’s chatgpt potentials in generating chatbot's dialogue for english as a foreign language learning.
\newblock {\em International Journal of Advanced Computer Science and Applications}, 14(6), 2023.

\bibitem{vrteachertrain21}
Özge Kelleci and N.~C. Aksoy.
\newblock Using game-based virtual classroom simulation in teacher training: User experience research.
\newblock {\em Simulation \& Gaming}, 52(2):204--225, 2021.

\end{thebibliography}
\end{document}